\documentclass[11pt]{article}
\usepackage{fullpage}
\usepackage{amsmath}
\usepackage{array}
\usepackage{epsfig}
\usepackage{graphicx}
\usepackage{amssymb}
\usepackage{color}
\usepackage{url}
\usepackage{authblk}
\usepackage{cite}

\newcommand{\diff}{\mathrm{d}}
\newcommand{\dg}{^{\mathrm{o}}}

\begin{document}
\title{Fast quantitative MRI as a nonlinear tomography problem}
\date{}
\author[$^{1,*}$]{Alessandro~Sbrizzi}
\author[$^1$]{Oscar~van~der~Heide}
\author[$^2$]{ Martijn~Cloos}
\author[$^1$]{Annette~van~der~Toorn}
\author[$^1$]{Hans~Hoogduin}
\author[$^1$]{Peter~R.~Luijten}
\author[$^1$]{Cornelis~A.~T.~van~den~Berg}
\affil[$^1$]{{\small Center for Image Sciences, University Medical Center, Utrecht}}
\affil[$^2$]{{\small Center for Advanced Imaging Innovation and Research (CAI$^2$R) Department of Radiology, New York University School of Medicine}}
\affil[$^*$]{{\small Corresponding author, contact:~\texttt{a.sbrizzi@umcutrecht.nl}}}
\maketitle

\begin{abstract}
Quantitative Magnetic Resonance Imaging (MRI) is based on a two-steps approach: estimation of the magnetic moments distribution inside the body, followed by a voxel-by-voxel  quantification of the human tissue properties. This splitting simplifies the computations but poses several constraints on the measurement process, limiting its efficiency. Here, we perform quantitative MRI as a one step process; signal localization and parameter quantification are simultaneously obtained by the solution of a large scale nonlinear inversion problem based on first-principles. As a consequence, the constraints on the measurement process can be relaxed and acquisition schemes that are time efficient and widely available in clinical MRI scanners can be employed. We show that the nonlinear tomography approach is applicable to MRI and returns human tissue maps  from very short experiments. \\ 
\textbf{Keywords:} MR-STAT, quantitative~MRI, nonlinear~tomography, MR~Fingerprinting, large~scale inversion.\\ \\
Article accepted for publication\footnote{\copyright~2017 This Manuscript version is made available under the CC-BY-NC-ND 4.0 license.} in \textit{Magnetic Resonance Imaging}, Volume 46 (2018), 56-63. DOI: 10.1016/j.mri.2017.10.015. 
\end{abstract}

\section{Introduction}
The possibility to store and process vast amounts of data at increasingly faster rates has boosted the application of numerical methods in physical sciences. Nowadays, solutions can be found to problems with hundred thousands or millions of unknowns \cite{wachter2006implementation,komodakis2015}. A representative example is seismic full waveform inversion \cite{fwi}; the underlying process is based on a  wave equation which is nonlinear in the spatially-dependent unknowns. The reconstruction  over 2D or 3D  regions of the Earth's interior is obtained by means of iterative  algorithms. It is even possible to estimate multiple parameters simultaneously, such as wave velocity, density, anisotropy and attenuation. \\
Analogously to seismic waveform inversion, quantitative magnetic resonance imaging (qMRI) aims at reconstructing several parameters which characterize the internal structure of the human tissue;  in particular, the proton density ($\rho$), the longitudinal ($T_1$) and transverse ($T_2$) relaxation rates, among others. \\
 One important difference between tomographic techniques and state of the art qMRI lies in their methodology. Quantitative MRI is built upon a two step approach. Firstly, each local contribution to the volumetric signal is estimated (signal localization), returning spatial maps of the transverse magnetic moment; this is usually achieved by applying a multi dimensional inverse Fourier transform to the data. Subsequently, the tissue parameters quantification is carried out for each location separately. The second step (parameter estimation) is thus obtained from a series of magnetization images by fitting  relatively simplistic signal models \cite{cheng2012} or by searching over a dictionary of complex signal fingerprints \cite{ma2013,cloos2016}.\\
This separation leads to a simplified computational process but with significant costs. In order to satisfy the stringent criteria for Fourier encoding, one has to assume that the signal evolution during the read-out only reflects the intended gradient encoding. Long single-shot read-outs generally violate this condition, leading to image artifacts, e.g., geometrical distortion and intra-voxel dephasing. To avoid such artifacts, most clinical MR sequences have been designed to manipulate the nuclear spins into a reproducible state, which allows multiple measurements to be aggregated into one coherent frequency representation of the
desired image ($k$-space). Consequently,  MRI scans can be relatively time consuming when compared to CT or PET exams. Additionally, due to the overly simplifying assumptions in the Fourier encoding-based signal model, system imperfections such as off-resonances and radiofrequency field inhomogeneity are not easily taken into account.\\
  MR Fingerprinting (MRF) \cite{ma2013} has shown a great potential to recover multi-parametric maps from unprecedented short acquisitions allowing strong aliasing artifacts to exist in each of the individual images. The RF excitation and gradient acquisition schemes need to be designed properly to ensure incoherence between the signal and the undersampling artifacts which are interpreted as zero-mean noise-like perturbations. Interleaved spiral \cite{ma2013}  and radial \cite{cloos2016} readout gradients are therefore preferred. These type of sequences are however, prone to gradient system imperfections such as eddy currents and thus require an additional sophisticated calibration of the hardware \cite{tan}. \\
In this work, we pose quantitative MRI as a nonlinear tomographic problem by directly utilizing the fundamental relationship between the time-varying signal and the laws of physics that describe the experiment. Thereby, we unify the traditionally disjoined processes of signal localization and parameter estimation into one process. The macroscopic ensemble of magnetic spins in the body is treated as a large-scale nonlinear dynamical system, which is probed by superimposing a train of radiofrequency (RF) excitations and gradient fields. The tissue properties are obtained by inversion of the underlying large scale nonlinear  model. We name this method MR-STAT, which stands for Magnetic Resonance Spin TomogrAphy in Time-domain. We show that quantitative parameter maps can be accurately reconstructed by employing nonlinear optimization algorithms and parallel computing infrastructures which do not necessarily rely on the Fourier decoding step for spatial localization. The data collection process can thus be liberated from the standard sequence design constraints and very short acquisitions (order of seconds) provides sufficient data for correct reconstructions. Although the time-domain formulation would in principle accommodate any read-out strategy, we show that established, experimentally robust cartesian gradient acquisition schemes can also be employed; a step which should facilitate the translation of the technique to clinical MRI systems. Finally, MR-STAT is also able to estimate the precision of the reconstructed multi-parametric maps;  another important step towards the clinical application of qMRI.
\section{Theory}
\subsection{The coupled space-time signal model}
\begin{figure}[!h]
\centering
\epsfig{file=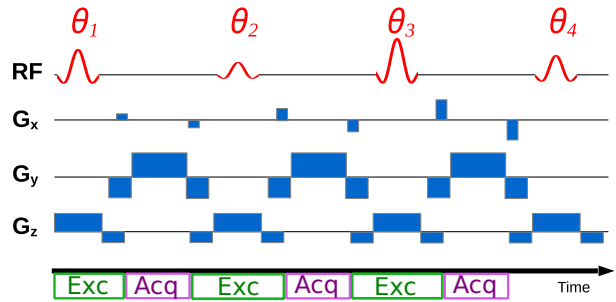, width=8cm}
\caption{\small{Fragment of an MR-STAT data acquisition sequence.  The spatially selective RF pulse is scaled by the tip angles $\theta_j$. $G_x$ and $G_y$ are encoding gradients. $G_z$ is the slice selective gradient. Note that the excitation (\textbf{Exc}) and acquisition (\textbf{Acq}) intervals follow one another without interruption, that is, the fixed echo and repetition times are the shortest possible.}}
\label{seq}
\end{figure}

The behavior of the space/time dependent magnetization vector, $\mathbf{m}(\mathbf{r},t)$ is determined by superimposed radiofrequency and gradient magnetic fields, respectively denoted by $b(t)$ and $\mathbf{G}(t)\cdot\mathbf{r}$. The response of the magnetic spins is also affected by the  $T_1(\mathbf{r})$ and $T_2(\mathbf{r})$ relaxation rates, which carry diagnostic information.  The relationship between all these quantities is given locally by the Bloch equation \cite{jaynes}:
\begin{equation}
 \frac{\diff}{\diff t}\mathbf{m}=\gamma\mathbf{b}\times\mathbf{m}-\mathbf{q}
\label{bloch}
\end{equation}
where 
\[
\mathbf{b}=
\left(
 \begin{array}{c}
\Re(b) \\
 \Im(b)\\
\mathbf{G}\cdot\mathbf{r}
\end{array} 
\right),\, 
\mathbf{q}=
\left(
\begin{array}{c}
\frac{m_x}{T_2}\\
\frac{m_y}{T_2}\\
\frac{m_z-1}{T_1}
\end{array}
\right),\,
\mathbf{m}(\mathbf{r},0)=
\left(
\begin{array}{c}
0\\
0\\
1
\end{array}
\right)               
\]
and $\gamma$ denotes the gyromagnetic ratio.\\
The signal, $s$, from a receiver coil is obtained from Faraday's law of induction \cite{brown2014}:
\begin{equation}
 s(t) = \int_V \rho(\mathbf{r}) m(\mathbf{r},t)\diff \mathbf{r}
\label{signal}
\end{equation}
where $\rho$ denotes the proton density of the tissue weighted by the spatially varying complex receive RF field $B_1^-$. $m$ is the transverse component of $\mathbf{m}$ and $V$ is the volume enclosing the spins which emit signal. 
\\
The first step in qMRI typically aims at reconstructing the spatially dependent magnetization state. This is achieved by designing the experiment such that the signal can be modeled as:
\begin{equation}
 s(t) = s(\mathbf{k}(t)) = \int_V\rho(\mathbf{r}) m^*(\mathbf{r})e^{-2\pi\imath\mathbf{k}(t)\cdot\mathbf{r}}\diff \mathbf{r}
\label{kspace}
\end{equation}
 where $m^*$ must be a  time-independent state of the magnetization and $\mathbf{k}$ represents the accumulating effect of the gradient fields. Note that the system response is decoupled into a space-only dependent component $\rho \,m^*$ and a Fourier encoding term $\exp(-2\pi\imath\mathbf{k}(t)\cdot\mathbf{r})$ which is independent from tissue parameters. The unknown term is thus $\rho \,m^*$. If Fourier transform requirements are fulfilled by the experimental settings, Inverse Fourier transform can be applied to the data to reconstruct $\rho\,m^*$, obtaining thus a magnetization image. This decoupled approach typically leads to either long measurement times ($m^*$ must be in the steady-states or in static equilibrium) or to large reconstruction artifacts if the Nyquist sampling criterion is not fulfilled \cite{ma2013}. In the subsequent step, model-fitting strategies based on the Bloch eq. (\ref{bloch}) can be applied to each voxel separately to recover the tissue parameters on a local level. In the MR fingerprinting case \cite{ma2013}, this is performed by an exhaustive search over a pre-computed dictionary of signals; a reconstruction strategy which although robust and straightforward,   is  undermined by the large dictionaries needed for high dimensional multi-parametric data. Furthermore, even a slight modification of a sequence requires an ad-hoc computation of the corresponding dictionary.   \\
Instead of relying on the standard decoupled Fourier model, we reconsider the coupled space-time equation, Eq. (\ref{signal}), and  solve it directly. Denoting by $d(t)$ the demodulated signal measured by the receiving coil of the MR scanner, the resulting tomographic approach is:
\begin{equation}
\begin{array}{c}
\text{Find the system's parameters, }\boldsymbol{\alpha}\text{, that minimize}\\ \\
\int_{\tau}|s(\boldsymbol{\alpha},t)- d(t)|^2\diff t\\ \\
\text{such that the Bloch eq. (1) and Faraday's law (2) hold.}
\end{array}
\label{mrstat}
\end{equation}
In the equation, $\tau$ denotes the union of temporal acquisition intervals and $\boldsymbol{\alpha}$ represents the unknown parameters over the whole region. Note that the reconstruction acts on the signal in time domain to directly derive the spatial distribution of the tissue's characteristics. In the MR-STAT framework, the link between temporal and spatial domain is still provided by the gradient fields, but now the $k$-space data set constitutes a non-trivial entanglement of spatial and
spin-dynamic information.\\
During an MR-STAT experiment, the magnetization is thus no longer expected to be in steady-states or equilibrium conditions but is free to evolve. Since there are no particular requirements on the state of the system, the excitation/acquisition scheme can be designed to boost the time-efficiency  and to minimize the impact of gradient hardware imperfections. In this work, we consider measurement schemes (sequences) where RF excitation pulses and acquisition intervals are contiguous, thus the repetition time $T_R$ and echo-time $T_E$ are kept as short as possible (see Fig. \ref{seq}); there are no dead times and the data collection rate is thus maximized. We choose to employ a so-called cartesian read-out scheme which is the standard acquisition modality due to its robustness with respect to hardware imperfections. \\
Since the reconstruction process no longer relies upon Fourier decoding, the underlying physical model can be easily expanded to include system imperfections such as off-resonance frequency, $\omega(\mathbf{r})$, and transmit RF fields heterogeneity, $B_1^+(\mathbf{r})$. These quantities enter the reconstruction problem (\ref{mrstat}) through the vector of applied magnetic field $\mathbf{b}$ in the Bloch equation (\ref{bloch}):
\[\mathbf{b}=(\Re(B_1^+b),\Im(B_1^+b),\mathbf{G}\cdot\mathbf{r}+\omega/\gamma)^T.\] 
Consequently, the extended set of unknowns in the MR-STAT equation (\ref{mrstat}) is 
\[\boldsymbol{\alpha} = (T_1,T_2,|\rho|,\angle{\rho},|B_1^+|,\omega).\]
 The MR-STAT reconstruction problem (Eq. (\ref{mrstat})) can be solved by a generic purpose derivative-based nonlinear minimization algorithm upon the discretization of the spatial and temporal domains. See the Methods section for the implementation details.
\subsection*{Spatial encoding, identifiability and precision estimates}
\begin{figure}[h!]
\centering
\epsfig{file=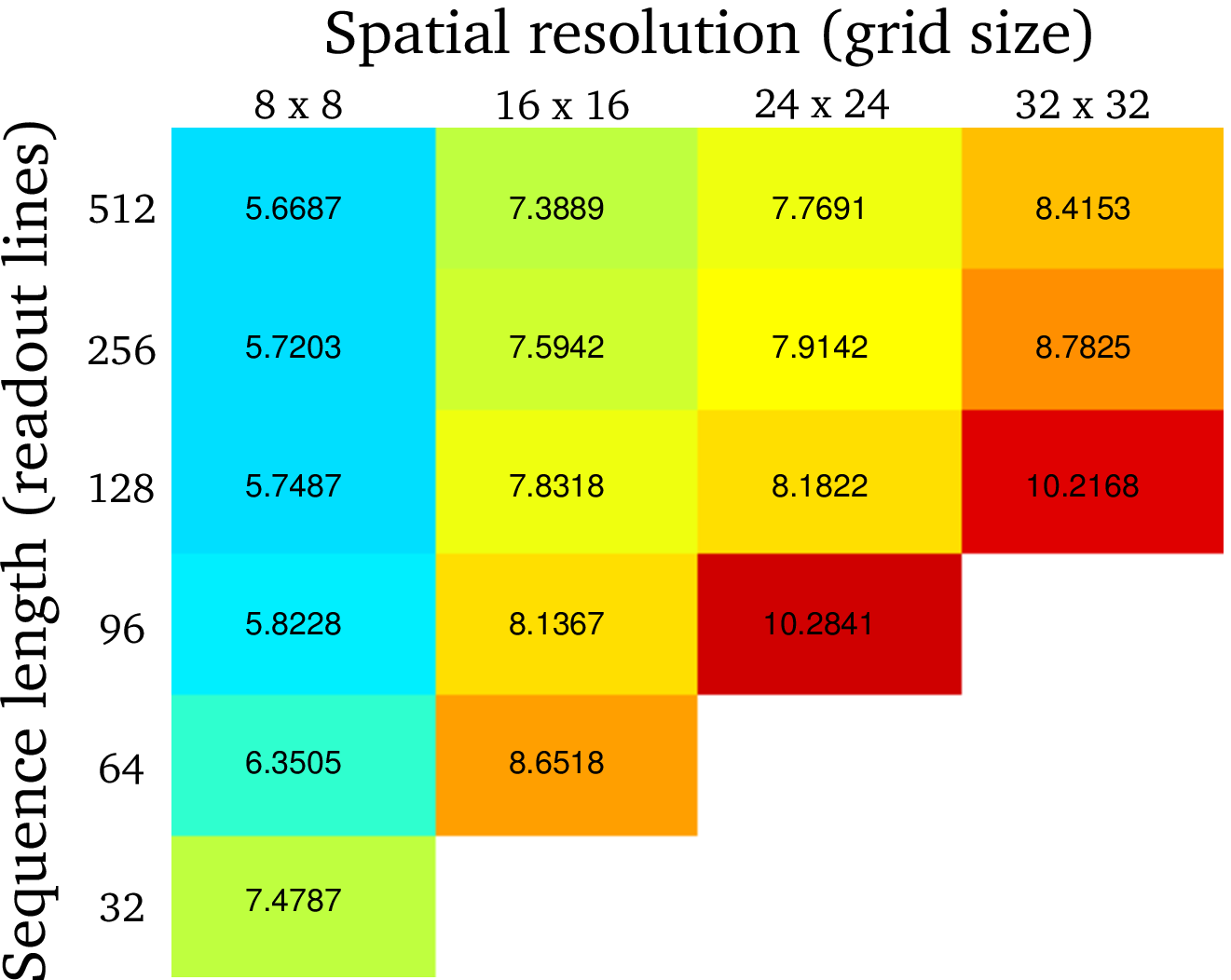, width=8cm}
\caption{\small{Noise amplification in the MR-STAT experiment when all six parameters $(T_1,T_2,|\rho|,\angle{\rho},|B_1^+|,\omega)$ are reconstructed. The numerical values inside the table refer to the $\log_{10}$ of the condition number of the covariance matrix $\mathbf{C}$. Large values mean large noise amplification. The condition number is reported as a function of the experiment length (numbers of readout lines) and spatial resolution (grid sizes) for a  small scale, homogeneous in-silico model. The number of samples per readout line is equal to the number of grid points along one dimension. An empty cell means that $\mathbf{C}$ is rank deficient (infinite condition number) and the problem can not be  solved.}}
\label{cond}
\end{figure}

 The  encoding capability of the MR-STAT approach can be derived by standard techniques in inversion theory. In particular, the identifiability of a system's parameters \cite{jacquez} is reflected by the covariance matrix $\mathbf{C}\equiv\eta^2( \mathbf{D}^T\mathbf{D})^{-1}$ where $\mathbf{D}$ is the Jacobian matrix of the model with respect to the parameters $\boldsymbol{\alpha}$ and $\eta$ is the  noise variance.\\
To minimize noise amplifications, $\mathbf{C}$ should have  a moderate condition number. This depends on both the acquisition length as well as the spatial resolution: for a fixed reconstruction grid, decreasing the sequence length leads to a more ill-conditioned matrix $\mathbf{C}$ and noise perturbations or model imperfections are thus amplified. In the extreme case that the sequence is too short, $\mathbf{C}$ becomes rank-deficient (infinitely large condition number) and the uniqueness of the solution is no longer guaranteed unless other regularization terms are introduced. This is analogous to reconstructions of undersampled $k$-space data in, for example, compressed sensing MRI \cite{cs,Doneva2010}. \\
To illustrate this theoretical analysis with a concrete example, we consider a homogeneous object with properties:
\[(T_1,T_2,|\rho|,\angle{\rho},|B_1^+|,\omega) =(0.833\text{[s]}, 0.083\text{[s]}, 1\text{[a.u.]}, 0\text{[rad]}, 1\text{[a.u.]}, 0\text{[Hz]})\]
 and construct $\mathbf{C}$ for varying spatial resolution and sequence length. The latter is expressed in terms of the number of readout lines in the sequence. The flip angles are randomly drawn from a normal distribution centered around 0 (see also the top of Fig. \ref{FA_signal}). The conditioning of the covariance matrix is reported in Fig. \ref{cond}. As expected, the longer the sequence, the lower the noise amplification. The number of unknowns increases with the grid size, leading to a larger scale problem requiring more data (longer sequences) to be fully determined and to be robust  to noise perturbations. 
When $\mathbf{C}$ has full rank, the MR-STAT problem is fully determined and the algorithm returns not only the parameter maps but also their spatially dependent standard deviations. The standard deviation of the $n$-th parameter is given by $\sigma_n\approx\sqrt{[\mathbf{C}]_{n,n}}$.
 Note the analogy between $\sigma_n$ and the so-called geometry factor (g-factor) in parallel imaging \cite{sense}.

\section{Methods}
\subsection{Implementation}
For reasons that will soon become clear, we split the vector of unknowns in two parts, namely: $\boldsymbol{\alpha} = (\boldsymbol{\rho},\boldsymbol{\beta})$ where $\boldsymbol{\beta}$ contains the spatial distribution of $(T_1,T_2,|B_1^+|,\omega)$. Given a demodulated dataset in the time domain, $d(t)$,  the reconstructed parameter maps, $(\boldsymbol{\rho}^{\text{rec}},\boldsymbol{\beta}^{\text{rec}})$, are obtained by  solving the following nonlinear least squares problem, which is derived upon the discretization of Eq. (\ref{mrstat}):
\begin{equation}
\begin{array}{c}
 (\boldsymbol{\rho}^{\text{rec}},\boldsymbol{\beta}^{\text{rec}}) = \arg\min_{\boldsymbol{\rho},\boldsymbol{\beta}} 
\sum\limits_{j =1}^J \left|d_j-\sum\limits_{r =1}^R \rho_r m_{j,r}(T_{1,r},T_{2,r},|B_{1,r}^+|,\omega_r)\Delta_s\right|^2\Delta_t, \\ \\
\text{such that Eq. (1) holds.}
\end{array}
\label{recon}
\end{equation}
 The first and second sum in the objective function approximate, respectively, the time and the volume integral from Eq. (\ref{mrstat})  and Eq. (\ref{signal}). $J$ is the total amount of acquired data samples, $R$ is the number of spatial grid points, $\Delta_s$ and $\Delta_t$ are, respectively, the space and time discretization intervals. Using matrix-vector notation, Eq. (\ref{recon}) can be written as:

\begin{equation}
\begin{array}{c}
 (\boldsymbol{\rho}^{\text{rec}},\boldsymbol{\beta}^{\text{rec}}) = \arg\min_{\boldsymbol{\rho},\boldsymbol{\beta}}\left\|\mathbf{d}-\mathbf{M}(\boldsymbol{\beta})\boldsymbol{\rho}\right\|^2 \\ \\
\text{such that Eq. (1) holds}
\end{array}
\label{matvec}
\end{equation}
where the matrix $\mathbf{M}(\boldsymbol{\beta})$ is given by  
\[[\mathbf{M}(\boldsymbol{\beta})]_{j,r}\equiv\Delta_s m_{j,r}(T_{1,r},T_{2,r},|B_{1,r}^+|,\omega_r).\]
 Since the reconstruction problem is nonlinearly dependent on  $\boldsymbol{\beta}$ and linearly dependent on $\boldsymbol{\rho}$, it can be solved by the variable projection method (VARPRO) \cite{golub}.  
Note that, if the vector $\boldsymbol{\beta}$ was a solution of Eq. (\ref{matvec}), then the parameters $\boldsymbol{\rho}$ could be found by solving a \textit{linear} least squares problem, whose solution is given by 
\begin{equation}
 \boldsymbol{\rho} = \mathbf{M}^{\dagger}(\boldsymbol{\beta})\mathbf{d}
\label{linls}
\end{equation}
where $\mathbf{M}^{\dagger}$ is the pseudo-inverse of $\mathbf{M}$. Substituting this back into Eq. (\ref{matvec}) we obtain the reduced functional:
\begin{equation}
 \boldsymbol{\beta}^* = \arg\min_{\boldsymbol{\beta}}\left\|\left[\mathbf{I}-\mathbf{M}(\boldsymbol{\beta})\mathbf{M}^{\dagger}(\boldsymbol{\beta})\right]\mathbf{d}\right\|^2.
\label{matvec1}
\end{equation}
Note that the linear parameter no longer plays a role in the equation.\\
VARPRO solves Eq. (\ref{matvec}) by first solving the reduced nonlinear problem in Eq. (\ref{matvec1}). The optimal linear parameters are eventually found by substitution into Eq. (\ref{linls}): $ \boldsymbol{\rho}^* = \mathbf{M}^{\dagger}(\boldsymbol{\beta}^*)\mathbf{d}$.\\
Solving Eq. (\ref{matvec1}) instead of Eq. (\ref{matvec}) results in a faster and robuster convergence for non-convex problems. Additionally, initial guesses for $\boldsymbol{\rho}$ are unnecessary. \\
The largest computational burden for solving Eq. (\ref{matvec1}) is given by the calculation of the derivatives of the system matrix $\mathbf{M}$ with respect to the nonlinear variables, that is: $\diff\mathbf{M}(\boldsymbol{\beta})/\diff\boldsymbol{\boldsymbol{\beta}}$. In this work, they are calculated by first order forward finite difference approximations.  We point out that the  VARPRO method has many applications and has even been used to solve different MR problems before \cite{boada1999,hernando2008,trzasko2013}. \\
The minimization problem is implemented in Matlab making use of the built-in trust region minimization algorithm and the VARPRO implementation given by \cite{oleary}. The Bloch equation simulator is implemented in C \cite{blochsim} and was adapted to include slice profile response, off-resonance effects and $B_1^+$ inhomogeneities.  The reconstruction is halted after 30 iterations or earlier if the maximum component of the gradient of the objective function is smaller than $10^{-6}$ (first order optimality measure). \\
Unless otherwise stated, the reconstruction algorithm is initialized with the following values:
\[
 (T_1,T_2,|B_1^+|,\omega)^{\text{start}}=(1.0 \text{[s]}, 0.1 \text{[s]}, 1.0 \text{[a.u.]}, 0.0 \text{[Hz]}).
\]
These values are uniform over the whole FOV. As explained, the (complex) proton density variable need not be initialized since it is reconstructed by solving a standard linear least squares problem. 

\subsubsection{Computational complexity and parallelization}
On the computation side, the  MR-STAT reconstruction problem for a 2D or 3D geometry at realistic spatial resolution is extremely demanding. Since all parameter maps are  reconstructed at once, the number of unknowns is vast. To illustrate: for a 2D acquisition of a $N_s \times N_s $ voxels grid, the number of unknowns is $N_s^2 \times 6$ since there are 6 parameters per voxel. Since $N_s \sim \mathcal{O}(10^2)$, the total number of unknowns is $ \mathcal{O}(10^5)$. As a consequence, the number of data points should also be $ \mathcal{O}(10^5)$. In addition,  the response of the system has also to be calculated  in the slice selective direction to correctly incorporate the effect of the slice profile. The reconstruction algorithm must calculate the response of the physical equations for $\mathcal{O}(10^5)$ voxels over $ \mathcal{O}(10^5)$ time points. \\
For the second and third reconstruction tests in this work (see below), we parallelize the computations in the following way:  suppose that we employ a Cartesian acquisition scheme with the read-out direction along the $y$-axis;  in this case, the signal, $s_j$, over the $j$-th read-out interval, $\tau_j$, is given by
\[
 s_j(t)  \propto \int_{X\times Y\times Z} m(\mathbf{r},t_j)e^{\frac{t_j-t}{T_2}}e^{\imath(t-t_j)\omega(\mathbf{r})}e^{-\imath\gamma G_y(t-t_j)y}\diff \mathbf{r} 
\]
where the 3D integration interval $X\times Y \times Z\subset\mathbb{R}^3$ contains all nuclear spins emitting a signal.
Given that for this kind of sequence, the duration of the read-out $\tau_j$ is only one millisecond or less, we can neglect the $T_2$ decay and the dephasing due to $\omega$. The signal equation becomes (we use the 1D $k$-space notation: $k_y\equiv\gamma/2\pi\int_{t_j}^t G_y(\tau)\diff\tau$):
\[
  s_j(k_y)  \approx \int_{X}\int_{Y}\int_{Z} m(x,y,z,t_j)e^{-i2\pi k_yy}\diff x\diff y\diff z
\]
and applying 1D Fourier Transform along the $y$ direction, $\mathcal{F}^y$:
\[
 \mathcal{F}^ys_j(\tilde{y}) \approx  \int_{X}\int_{Z} m(x,\tilde{y},z,t_j)\diff x\diff z
\label{sig1}
\]
$\mathcal{F}^ys_j(\tilde{y})$ represents the signal generated at time $t_j$ by the nuclear spins located in the 2D interval $X\times Z$ at the $y$-coordinate given by $\tilde{y}$. The signal from spins with different $y$-coordinates does not contribute to $\mathcal{F}^ys_j(\tilde{y})$. In other words: the MR-STAT reconstruction problem can be decomposed into many independent subproblems, each one corresponding to a given coordinate $\tilde{y}_n$ with $n=1,\dots,N_s$. Parallelization is thus carried out by assigning each subproblem to a different computing core. The reconstruction time  is defined as the longest runtime amongst all jobs.\\
The whole code is compiled as a Linux stand-alone executable and deployed to the High Performance Computing cluster of the UMC Utrecht by linking it to the corresponding Matlab run-time library.
\subsection{Reconstructions}
To demonstrate the design flexibility of MR-STAT, we employ several types of acquisition schemes: one where the tip angles are randomly drawn from a normal distribution (Fig. \ref{FA_signal}); one which follows a sinusoidal pattern where each lobe is weighted by a randomly chosen value (Fig. \ref{invivo_sin}-Top) and one with piecewise constant excitations (Fig. \ref{invivo_gre}-Top). For the latter RF-train, each constant tip angle section is preceded and followed by a half-angle pulse acting, respectively, as excitation and tip-back pulses. All the sequences start with a 180$\dg$ inversion pulse. Each read-out interval is centered between excitations and all gradients are balanced, thus a single isochromat accurately represents the dynamics of a voxel.
\subsubsection{\textit{In silica} low resolution reconstruction}
A simple 2D object made of three homogeneous compartments is reconstructed on a $32\times 32$ grid (See Fig. \ref{toy}). The $T_1$ and $T_2$ rates for the three compartments A, B, and C correspond to cerebrospinal fluid (CSF), gray and white matter values, respectively. In this case, the off-resonance and transmit RF maps were set to $\omega=0$ Hz and $B_1^+=1$ [a.u.], respectively. A random RF excitation train is applied analogously to the one shown in Fig. \ref{FA_signal}. Two-hundred and fifty-six RF pulses are interleaved with a 2D Cartesian read-out gradient scheme consisting of 32 phase encoding steps which are repeated 8 times. The resulting sequence duration is 1.2 seconds. Gaussian noise is superimposed to the time-domain signal such that $\|\text{noise}\|/\|\text{signal}\|=0.01$. 
\subsubsection{\textit{In-silica} high resolution reconstruction}
\begin{figure}[!t]
\centering
\epsfig{file=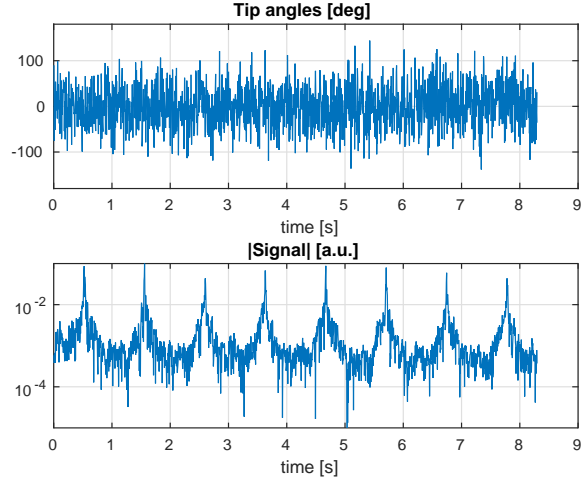, width=9cm}
\caption{\small{Tip angles and time-domain signal for the MR-STAT sequence applied to the in-silico simulated head experiment at 3T.}}
\label{FA_signal}
\end{figure}

\begin{figure}[h!]
\centering
\epsfig{file=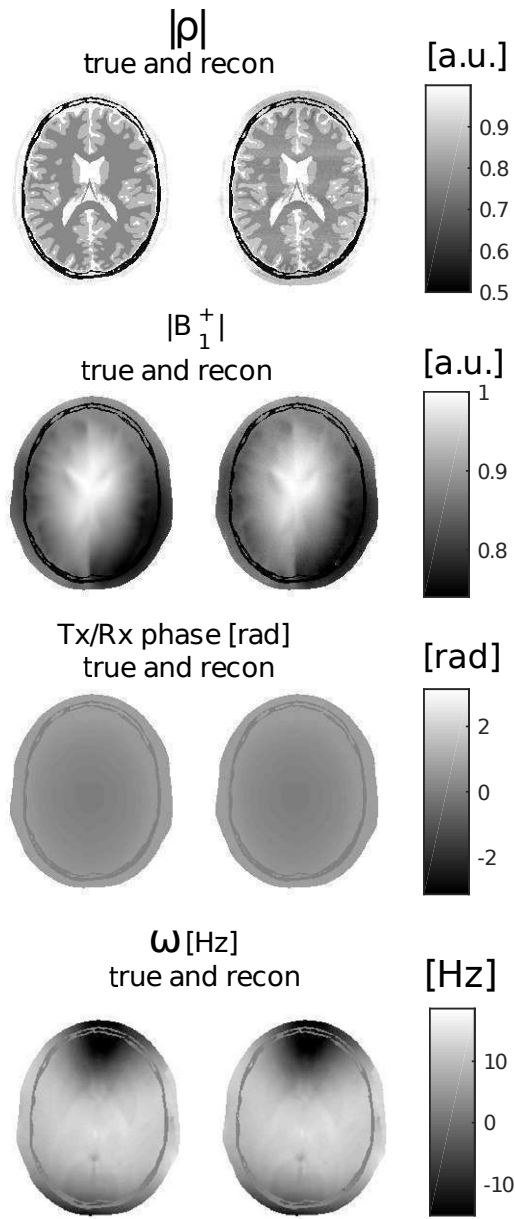, width=7cm}
\caption{\small{True and reconstructed maps of proton density, transceive phase, $|B_1^+|$ and $\omega$.}}
\label{other_recon}
\end{figure}

The central slice of a numerical human brain model \cite{brainweb} is used to create a synthetic MR-STAT data set. The reconstructed in-plane resolution is 1mm$\times1$mm which corresponds to a $216\times 216$ voxels matrix. The tissue parameters for the biological components are given in Table \ref{tab:paramsrec}. 
The amplitude and phase maps of the  transmit RF field are obtained from a numerical electromagnetic simulation of a 3T headcoil driven in quadrature. Without loss of generality, a uniform receive sensitivity is assumed in this example. The off-resonance map is taken from \cite{B0map} and  is scaled to fit the range of $[-15, 15]$ Hz in the head (see the bottom of Fig.  \ref{other_recon}). 
For the acquisition, a Cartesian trajectory is used. The duration of each read out is 0.86 ms with a 4 $\mu$s dwell time per sample. The read out lines ($k_y$ direction) cover the 2D $k$-space in ascending order, starting with the smallest negative values of $k_x$ and repeating this pattern for the equivalent of 8 full $k$-space coverages. In total, $1728$ lines are acquired in 8.3 seconds resulting in approximately $3.7\cdot 10^5$ time data points. The random tip angles sequence is shown at the top of Fig. \ref{FA_signal}.\\
A Gaussian shaped RF pulse and a slice selective gradient waveform along the $z$ axis are applied. The  RF pulse is 1 ms long and is defined on a $0.1$ ms dwell time step. The slice profile variation throughout the sequence is taken into account by discretizing the spatial domain in the slice-selective direction by 50 points and integrating the magnetization response for each point. This integration is applied to both the forward (signal simulation) and backward (reconstruction) steps. Gaussian noise is superimposed to the time-domain signal such that $\|\text{noise}\|_2/\|\text{signal}\|=0.01$. The resulting time-domain signal is shown at the bottom of Fig. \ref{FA_signal}.\\
The parameter $\omega$ is initialized by applying a median filter to the true off-resonance map. In experimental practice, this dataset could be generated with a fast $B_0$ calibration scan. The other parameters are initialized with the same values as reported in the Implementation subsection.
\subsubsection{\textit{In-vivo} experimental demonstration at 3.0 Tesla}
\begin{figure}[!t]
\centering
\epsfig{file=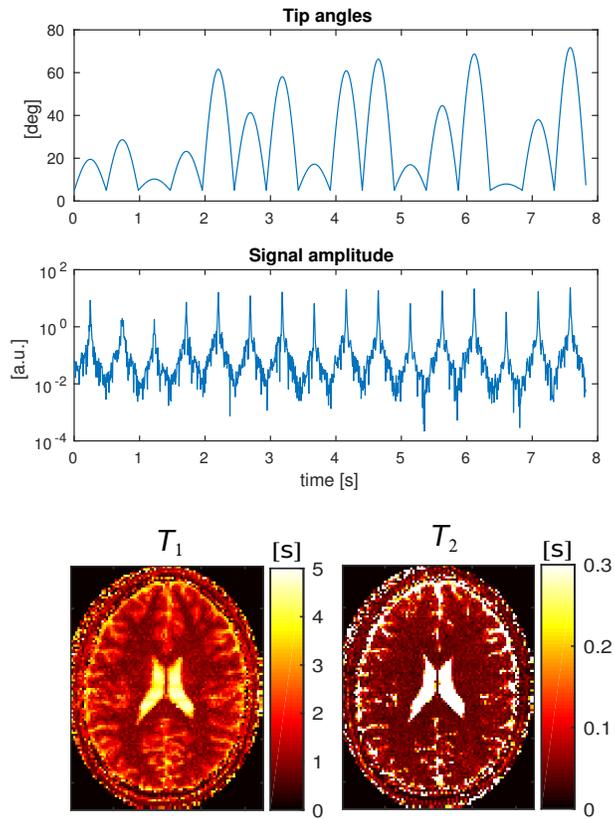, width=8cm}
\caption{\small{In-vivo experimental validation of MR-STAT for a sinusoidal RF train sweep. From top to bottom: the flip angle train, the recorded signal and the reconstructed parameter maps. }}
\label{invivo_sin}
\end{figure}

\begin{figure}[!t]
\centering
\epsfig{file=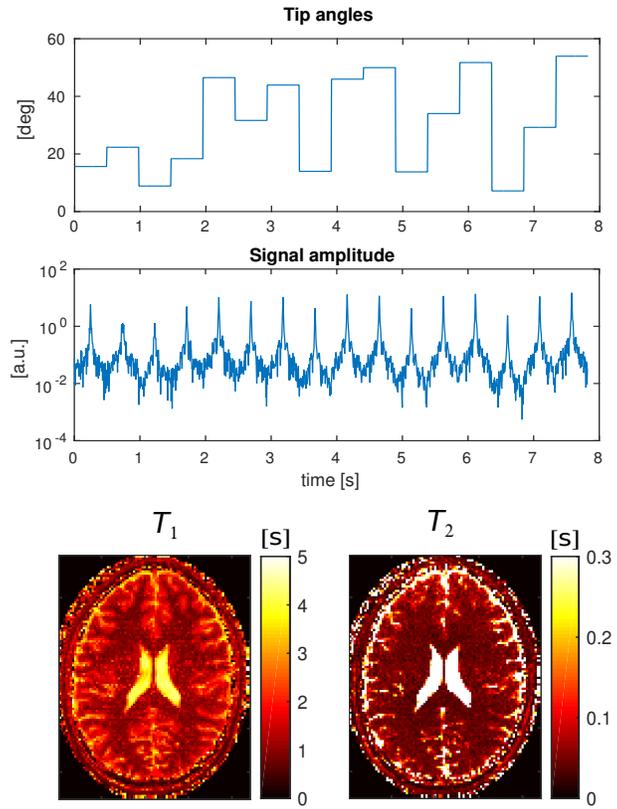, width=8cm}
\caption{\small{In-vivo experimental validation of MR-STAT for a piecewise constant tip angle excitation. From top to bottom: the flip angle train, the recorded signal and the reconstructed parameter maps.}}
\label{invivo_gre}
\end{figure}

Finally, MR-STAT is implemented on a 3T whole-body MR system (Philips-Ingenia). A single slice is acquired for a brain of a healthy volunteer with a 15 channel receive head-coil. Written informed consent from the volunteer participating in this experiment was obtained.\\
We employ two different sequences. The first RF train (Fig. \ref{invivo_sin}, top) consists of 16 sinusoidal sweeps. Each lobe corresponds to a $k$-space filling and is randomly scaled to achieve maximum amplitude levels in the range $5\dg\leq\theta\leq 75\dg$.  \\
The second RF train (Fig. \ref{invivo_gre}, top) consists of piecewise constant flip angles, whose values are drawn from a uniform distribution in the range $[5\dg,60\dg]$. Each of the 16 $k$-space fillings is thus characterized by the same tip angle excitation.  In addition, a half-angle pre-pulse and a half angle tip-back pulse are applied, respectively, before and after each segment.\\
In both sequences, the excitation phases alternate between $0\dg$ and $180\dg$.  A Gaussian shaped RF pulse with duration 0.81 ms and a slice selective gradient are employed to achieve a 3mm slice thickness. The shortest possible values for $T_E$ and $T_R$ are chosen, namely $(T_E,T_R)=(2.78,5.56)$ ms. The sequences are preceded by an adiabatic inversion pulse. The sequence parameters are converted into MATLAB format and imported in the reconstruction software. Analogously to the synthetic case, the slice profile variation across the sequence is included in the model by simulating the RF pulses on a 15 $\mu$s grid and taking 11 samples along the slice direction.  As starting values for $\omega$ we choose 0 Hz everywhere.\\
The spatial resolution is $1.8\times 1.8$ mm$^2$ and the scan time is 7.8 seconds. The measured signals are shown in Figures \ref{invivo_sin} and \ref{invivo_gre}.\\
  In these two tests, we reconstruct $T_1$ and $T_2$ value and we treat the other parameters as nuisance variables, that is, they are considered unknown but their estimation is not required to be precise.
 
 \section{Results}
\subsection{\textit{In silica} low-resolution reconstruction}
\begin{figure}[!t]
\centering
\epsfig{file=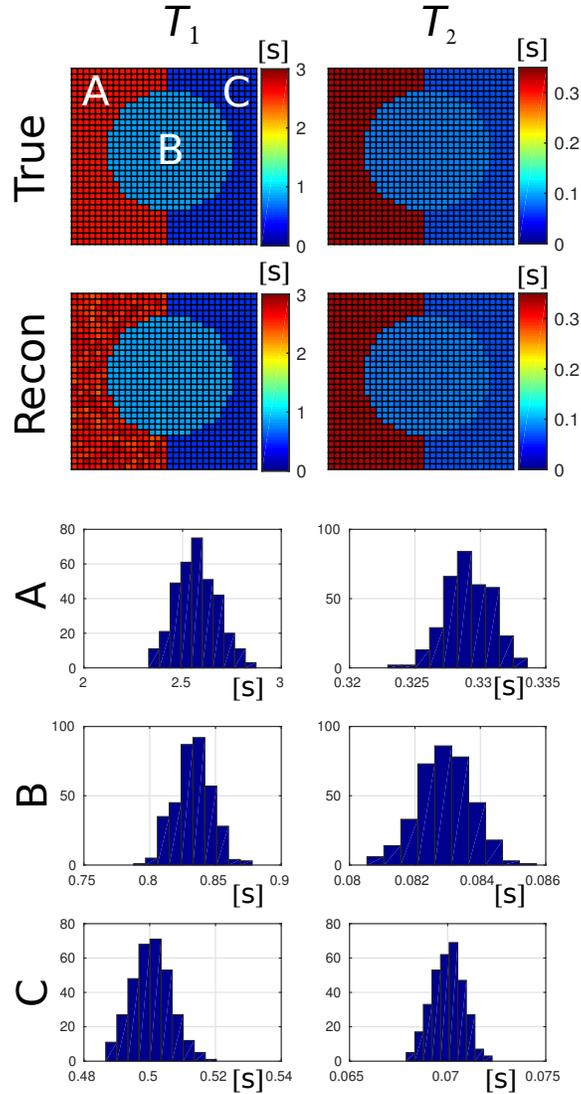, width=8cm}
\caption{\small{Precision estimate test. A simple 2D object (top row) undergoes a simulated MR-STAT acquisition and reconstruction. The reconstructed $T_1$ (left) and $T_2$ (right)  maps are shown on the second row. The histogram plots report the distribution of the reconstructed  values over each compartment A, B and C. The standard deviations of these distributions are reported in Table \ref{toy_table} and show great similarity with the estimated standard deviation values.}}
\label{toy}
\end{figure}

\begin{figure}[!h]
\centering
\epsfig{file=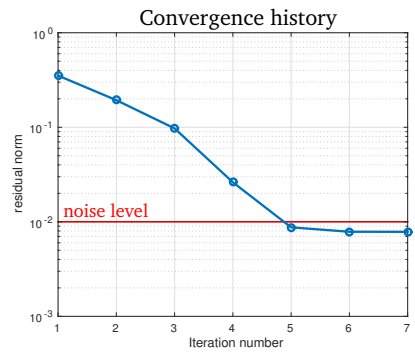, width=6cm}
\caption{\small{Convergence curve of the MR-STAT reconstruction algorithm for the precision estimation test (see also Fig. \ref{toy}). The relative residual norm (data-model misfit normalized on the norm of the data) is reported as a function of the iteration number. Note that the algorithm eventually converges to the thermal noise level.}}
\label{history}
\end{figure}

Fig. \ref{toy} illustrates  the application of MR-STAT  to the small scale reconstruction test. The distribution of reconstructed values from each compartment are reported in the histogram plots. The  standard deviations  as estimated from the covariance matrix $\mathbf{C}$ are averaged over each compartment and are reported in Table \ref{toy_table}. In the same Table, also the true standard deviations obtained from the reconstructed values are reported. These are calculated as  
\[\sqrt{\frac{1}{K-1}\sum_{k=1}^{K}|T_i^{\text{recon}}-T_i^{\text{true}}|^2},\quad i=1,2\]
where $K$ is the number of voxels in a given compartment. From   Table 1 it is clear that not only the $T_1$ and $T_2$ values are accurately reconstructed (as shown in Fig. \ref{toy}), but also the estimated and truly obtained precision levels are very similar.\\
The convergence curve for the reconstruction algorithm is reported in Fig. \ref{history} and displays the relative residual norm as a function of the iteration number, that is, the model-data misfit normalized on the norm of the data:
\[
 \text{relative residual norm} =\frac{\left\|\mathbf{d}-\mathbf{M}(\boldsymbol{\beta})\boldsymbol{\rho}\right\|}{\|\mathbf{d}\|}.
\]
 The data-model misfit eventually reaches the noise level after 5 iterations and the algorithm halts soon afterwards.
\subsection{\textit{In-silica} high resolution reconstruction}
\begin{figure}[h!]
\centering
\epsfig{file=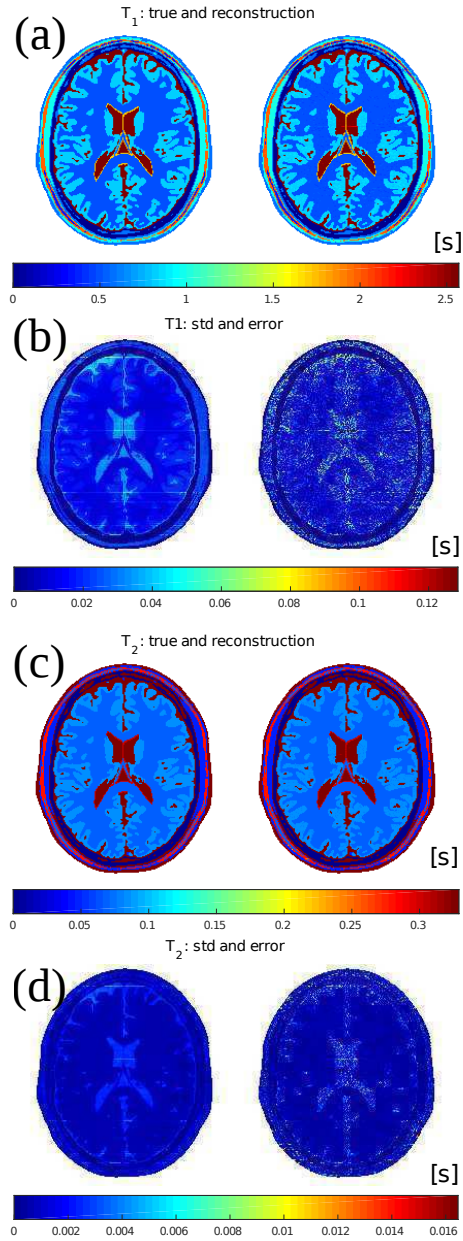, width=6cm}
\caption{\small{$T_1$ and $T_2$ maps for the synthetic MR-STAT acquisition and reconstruction. (a) and (c): true and reconstructed maps. (b) and (d): standard deviation maps estimated by MR-STAT and the error in the reconstructions ($|T_1^{\text{true}}-T_1^{\mathrm{recon}}|$ and $|T_2^{\text{true}}-T_2^{\mathrm{recon}}|$).}}
\label{T1T2}
\end{figure}

Beside $T_1$, $T_2$ and $\rho$, also the transmit field profile and off-resonance map are reconstructed; they are displayed in Figures \ref{T1T2} and \ref{other_recon}. They closely agree with the true values. In Table \ref{tab:paramsrec}, the mean values and corresponding variations over each tissue type are reported and show high precision.\\ 
The root-mean-squared-errors (RMSE) for the $B_1^+$ and $\omega$ maps are also very small, namely:
\[
 \text{RMSE}(|B_1^+|) = 0.0043 \text{ [a.u.]},\quad\text{RMSE}(\omega) = 0.12 \text{ [Hz]}.
\]
The reconstruction time  is about 90 minutes. The median number of performed iterations  as calculated over all parallel reconstruction processes is 13.\\
The standard deviations estimated by MR-STAT  for $T_1$ and $T_2$ are shown, respectively, in Fig. \ref{T1T2}(b) and Fig. \ref{T1T2}(d). For comparison, the actual error maps, respectively defined as $|T_1-T_1^{\mathrm{recon}}|$ and $|T_2-T_2^{\mathrm{recon}}|$, are also reported and they show clear similarities.  

\subsection{\textit{In-vivo} experimental demonstration at 3.0 Tesla}
 The obtained  $T_1$ and $T_2$ maps are shown at the bottom of Fig. \ref{invivo_sin} and Fig. \ref{invivo_gre}. The reconstruction algorithm was halted after 12 iterations since the solution did not significantly improved during the last few iterations. The computation time was about 12 minutes for both datasets.

\section{Discussion}
Traditional quantitative MR methods are typically performed in two steps; first a series of images is reconstructed, then the quantitative parameters are estimated from these images on a voxel-by-voxel basis. The recently introduced MRF method \cite{ma2013} works along similar lines, but shifts the focus away from the signal localization process and onto the temporal dynamics of the spin-system. Although MRF still adheres to the traditional two step procedure, it sacrifices accurate signal triangulation in favour of a high sampling rate. The resulting undersampling artifacts in each image are treated as a large, zero-mean, noise-like process, thus the signal model includes   a substantial pseudo-stochastic component.   MR-STAT relies instead on a fully deterministic strategy by employing a coupled space-time model that encapsulates the entire MR experiment. Consequently, the model accuracy is drastically enhanced and the brute-force exhaustive search is replaced by iterative minimization methods which exploit the structure of the underlying dynamics. The MR-STAT approach aims thus at a better utilization of the information carried by the data and to the elimination of the dictionary search, which is notoriously hindered by the curse of dimensionality. Another important benefit of taking this route is that it provides deep insights into the important aspect of error estimation. The availability of standard deviation maps is a valuable tool for quality monitoring; a fundamental aspect for the clinical application of quantitative MRI. \\
It is important to realize that the gradient trajectory used in MR-STAT does not necessarily relate directly to the spatial resolution. The $k$-space in MR-STAT is not a spatial frequency domain, as is the case in standard MRI acquisition approaches. Although some demonstrations shown here still use a one dimensional Fourier transform along the read-out direction for parallelization, the MR-STAT formalism can, in principle, remove the explicit Fourier relationship between the time and image domain in its entirety. This will be beneficial in the case of non-cartesian trajectories such as  radial and spiral or for non-linear gradient field systems \cite{patlock}. As we move more and more along this direction, it may be better to think of trajectories in gradient space than in an actual $k$-space. Inversion theory provides tools to generalize the concept of  encoding capability for transient-states sequences when time and space dependence are implicitly entangled in the signal and results from Fourier theory are no longer applicable. \\
 The primary cost of the MR-STAT approach is that all quantitative  parameters must be estimated at once, which leads to a formidable inversion problem. We have however been able to reconstruct multi-parametric maps using a high performance computing facility within a reasonable computation time. The experimental design is more flexible since neither steady-states or static equilibrium conditions are  needed nor the incoherence between undersampling artifacts and true signal; this  allows for very short acquisitions (few seconds for a 2D slice) based upon experimentally reliable cartesian read-out schemes. In one of the experiments (see Fig. \ref{invivo_gre}), we employed a step-wise flip-angle scheme combined with a standard bSSFP sequence, which is a widely available protocol on regular MR systems and does not require major adaptations on the acquisition. Also on the reconstruction side, flexibility is guaranteed by the inverse approach of MR-STAT; any changes made by the operator at the console during the exam can be easily accommodated in the reconstruction. \\
MR-STAT has been developed upon the philosophy that   scanner time is much more expensive than  computing time. We believe that this gap will keep growing in the future as computing power and algorithmic acceleration constantly increase.  The current trends in bio-informatics and genomics show that local computing clusters or  cloud computing on remote servers are becoming increasingly available in a hospital setting. The moderate investment in terms of the required computing infrastructure is highly profitable given the potential of MR-STAT for improving cost-effectiveness and patient comfort due to the reduced scan times and more simple workflows.  \\
This study has focused on the computational and experimental proof-of-principle of MR-STAT. There is room  left to study and optimize the accuracy,  precision and speed of this framework. For instance, regularization techniques could be applied to reduce the noise amplification in the in-vivo measurements.  Other techniques that could enhance MR-STAT are parallel imaging \cite{smash,sense,grappa} and compressed sensing \cite{cs,Doneva2010}. The availability of multiple independent receivers and sparsity regularization terms can dramatically improve the triangulation of the signal origins thus greatly improving the conditioning of the comprehensive optimization problem. In general, optimum experiment design techniques \cite{box,xiao} could be applied to maximize the differentiation between signal evolutions and possibly enhance the rate of convergence while maintaining short acquisition times. \\
With this work, we intended to prove that quantitative MRI can be treated as a nonlinear tomographic problem and therefore large scale nonlinear optimization techniques can be successfully applied. We hope that that this manuscript will inspire researchers from other fields, to try and apply their experience and knowledge in the area of large scale
inversion problems to the qMRI and medical imaging in general.
\section{Conclusion}
A new framework for multi-parametric quantitative MRI, called MR-STAT, has been presented. Signal localization and parameter estimation are solved simultaneously by inverting a coupled space-time model from time domain data. This is obtained by established large scale nonlinear inversion techniques running on a high performance computing facility. The measurement efficiency is boosted by the elimination of dead times and traditional assumptions that inject artifacts into standard reconstruction approaches are circumvented. Moreover, this new formalism provides insights into the precision estimation of fast quantitative MRI.
\section{Acknowledgment}
Part of this work was funded by the Dutch Technology Foundation (NWO-STW), grant number 14125.\\
The authors are grateful to Dr. Tristan van Leeuwen, Prof. Jeannot Trampert and Dr. Ivan Vasconcelos for fruitful discussions and to Mrs Ying Lai Green for proofreading the manuscript. 

\bibliography{bibfile_v6.1}{}

\begin{thebibliography}{10}

\bibitem{wachter2006implementation}
Andreas W{\"a}chter and Lorenz~T Biegler.
\newblock On the implementation of an interior-point filter line-search
  algorithm for large-scale nonlinear programming.
\newblock {\em Mathematical programming}, 106(1):25--57, 2006.

\bibitem{komodakis2015}
Nikos Komodakis and Jean-Christophe Pesquet.
\newblock Playing with duality: {An} overview of recent primal-dual approaches
  for solving large-scale optimization problems.
\newblock {\em IEEE Signal Processing Magazine}, 32(6):31--54, 2015.

\bibitem{fwi}
Jean Virieux and St{\'e}phane Operto.
\newblock An overview of full-waveform inversion in exploration geophysics.
\newblock {\em Geophysics}, 74(6):WCC1--WCC26, 2009.

\bibitem{cheng2012}
Hai-Ling Margaret~Cheng, Nikola Stikov, Nilesh~R Ghugre, and Graham~A Wright.
\newblock Practical medical applications of quantitative {MR} relaxometry.
\newblock {\em Journal of Magnetic Resonance Imaging}, 36(4):805--824, 2012.

\bibitem{ma2013}
Dan Ma, Vikas Gulani, Nicole Seiberlich, Kecheng Liu, Jeffrey~L Sunshine,
  Jeffrey~L Duerk, and Mark~A Griswold.
\newblock Magnetic resonance fingerprinting.
\newblock {\em Nature}, 495(7440):187--192, 2013.

\bibitem{cloos2016}
Martijn~A Cloos, Florian Knoll, Tiejun Zhao, Kai~T Block, Mary Bruno, Graham~C
  Wiggins, and Daniel~K Sodickson.
\newblock Multiparametric imaging with heterogeneous radiofrequency fields.
\newblock {\em Nature communications}, 7, 2016.

\bibitem{tan}
Hao Tan and Craig~H Meyer.
\newblock Estimation of k-space trajectories in spiral {MRI}.
\newblock {\em Magnetic resonance in medicine}, 61(6):1396--1404, 2009.

\bibitem{jaynes}
ET~Jaynes.
\newblock Matrix treatment of nuclear induction.
\newblock {\em Physical Review}, 98(4):1099, 1955.

\bibitem{brown2014}
Robert~W Brown, Y-C~Norman Cheng, E~Mark Haacke, Michael~R Thompson, and Ramesh
  Venkatesan.
\newblock {\em Magnetic resonance imaging: physical principles and sequence
  design}.
\newblock John Wiley \& Sons, 2014.

\bibitem{jacquez}
John~A Jacquez and Peter Greif.
\newblock Numerical parameter identifiability and estimability: Integrating
  identifiability, estimability, and optimal sampling design.
\newblock {\em Mathematical Biosciences}, 77(1-2):201--227, 1985.

\bibitem{cs}
Michael Lustig, David Donoho, and John~M Pauly.
\newblock Sparse mri: {The} application of compressed sensing for rapid {MR}
  imaging.
\newblock {\em Magnetic resonance in medicine}, 58(6):1182--1195, 2007.

\bibitem{Doneva2010}
Mariya Doneva, Peter B{\"o}rnert, Holger Eggers, Christian Stehning, Julien
  S{\'e}n{\'e}gas, and Alfred Mertins.
\newblock Compressed sensing reconstruction for magnetic resonance parameter
  mapping.
\newblock {\em Magnetic Resonance in Medicine}, 64(4):1114--1120, 2010.

\bibitem{sense}
Klaas~P Pruessmann, Markus Weiger, Markus~B Scheidegger, Peter Boesiger, et~al.
\newblock {SENSE}: sensitivity encoding for fast {MRI}.
\newblock {\em Magnetic resonance in medicine}, 42(5):952--962, 1999.

\bibitem{golub}
Gene Golub and Victor Pereyra.
\newblock Separable nonlinear least squares: the variable projection method and
  its applications.
\newblock {\em Inverse problems}, 19(2):R1, 2003.

\bibitem{boada1999}
Fernando Boada, Zhi-Pei Liang, and E~Mark Haacke.
\newblock Improved parametric reconstruction using variable projection
  optimization.
\newblock {\em Inverse problems}, 14(1):19, 1998.

\bibitem{hernando2008}
Diego Hernando, JP~Haldar, BP~Sutton, J~Ma, P~Kellman, and Z-P Liang.
\newblock Joint estimation of water/fat images and field inhomogeneity map.
\newblock {\em Magnetic resonance in medicine}, 59(3):571--580, 2008.

\bibitem{trzasko2013}
Joshua~D Trzasko, Petrice~M Mostardi, Stephen~J Riederer, and Armando Manduca.
\newblock Estimating {T1} from multichannel variable flip angle {SPGR}
  sequences.
\newblock {\em Magnetic resonance in medicine}, 69(6):1787--1794, 2013.

\bibitem{oleary}
Dianne~P O’leary and Bert~W Rust.
\newblock Variable projection for nonlinear least squares problems.
\newblock {\em Computational Optimization and Applications}, pages 1--15, 2013.

\bibitem{blochsim}
\url{http://mrsrl.stanford.edu/~brian/blochsim}.

\bibitem{brainweb}
Berengere Aubert-Broche, Alan~C Evans, and Louis Collins.
\newblock A new improved version of the realistic digital brain phantom.
\newblock {\em NeuroImage}, 32(1):138--145, 2006.

\bibitem{B0map}
Website of the {National} {Alliance} for {Medical} {Image} {Computing}.
  {Available}:
  \url{http://wiki.na-mic.org/Wiki/index.php/Projects:QuantitativeSusceptibilityMapping}.

\bibitem{patlock}
Juergen Hennig, Anna~Masako Welz, Gerrit Schultz, Jan Korvink, Zhenyu Liu,
  Oliver Speck, and Maxim Zaitsev.
\newblock Parallel imaging in non-bijective, curvilinear magnetic field
  gradients: a concept study.
\newblock {\em Magnetic Resonance Materials in Physics, Biology and Medicine},
  21(1):5--14, 2008.

\bibitem{smash}
Daniel~K Sodickson and Warren~J Manning.
\newblock Simultaneous acquisition of spatial harmonics ({SMASH}): fast imaging
  with radiofrequency coil arrays.
\newblock {\em Magnetic Resonance in Medicine}, 38(4):591--603, 1997.

\bibitem{grappa}
Mark~A Griswold, Peter~M Jakob, Robin~M Heidemann, Mathias Nittka, Vladimir
  Jellus, Jianmin Wang, Berthold Kiefer, and Axel Haase.
\newblock Generalized autocalibrating partially parallel acquisitions
  ({GRAPPA}).
\newblock {\em Magnetic resonance in medicine}, 47(6):1202--1210, 2002.

\bibitem{box}
G~Eo~P Box and HL~Lucas.
\newblock Design of experiments in non-linear situations.
\newblock {\em Biometrika}, 46(1/2):77--90, 1959.

\bibitem{xiao}
Z~Xiao and A~Vien.
\newblock Experimental designs for precise parameter estimation for non-linear
  models.
\newblock {\em Minerals Engineering}, 17(3):431--436, 2004.

\end{thebibliography}
\bibliographystyle{unsrt}
\newpage
\section*{Tables}

\begin{table}[!h]
\caption{\small{Precision estimation test. Estimated standard deviations per compartment as derived from the covariance matrix $\mathbf{C}$ versus the true standard deviation calculated after the reconstruction. The estimated precision levels are very close to the obtained ones.}}
\centering
\begin{tabular}{l|rr|rr}
 & \multicolumn{2}{c|}{$T_1$}  & \multicolumn{2}{c}{$T_2$} \\
 Compartment &   $\sqrt{[\mathbf{C}]_{n,n}}$ &std of recon & $\sqrt{[\mathbf{C}]_{n,n}}$ &std of recon \\
\hline
A (CSF)& 112.8 [ms]& 114.1  [ms]&2.0 [ms]& 1.8 [ms]\\
B (Gray m.) & 16.1 [ms]&  14.2  [ms]& 0.9  [ms]& 0.8 [ms]\\
C (White m.)  &  6.6 [ms]&  5.8 [ms]&  0.9   [ms]& 0.8 [ms]\\
\hline
\end{tabular}
\label{toy_table}
\end{table}

\begin{table}[!h]
\caption{\small{True and Mean values of the reconstructed relaxation times per tissue type. The standard deviation of the reconstructed values for each tissue type is reported in brackets.}}
\centering
\begin{tabular}{l|rrl|rrl}
 & \multicolumn{3}{c|}{$T_1$ [ms] }  & \multicolumn{3}{c}{$T_2$ [ms] } \\
 &  true & recon &(std) & true &recon &(std) \\
\hline
CSF & 2569&2565.7 &($\pm$38.9)&329&329.1 &($\pm$2.8)\\
Gray m.  & 833&833.4 &($\pm$18.9)   &83 & 83.0 &($\pm$0.8)\\
White m.   & 500&500.9 &($\pm$12.2)  & 70 & 70.0 &($\pm$0.6)\\
  Fat      &  350&352.2  &($\pm$8.9)  & 70 &70.0 &($\pm$0.5) \\
Muscle &1000&1000.6 &($\pm$31.0)  &47  & 47.0 &($\pm$0.6)\\
Skin & 569&570.1  &($\pm$7.7) & 329&328.3 &($\pm$4.0)\\
Blood   & 1700&1699.3 &($\pm$21.7)  & 300 & 299.6 &($\pm$2.5)\\
Dura   & 2000&2001.1 &($\pm$41.1)  &280  &279.2 &($\pm$5.2)\\
\hline
\end{tabular}
\label{tab:paramsrec}
\end{table} 

\newpage
\section*{Figures}

\end{document}